\documentclass[prr,superscriptaddress,floatfix,twocolumn,showpacs]{revtex4-2}
\usepackage{babel}
\usepackage{fmtcount}
\usepackage{verbatim}
\usepackage{amsmath}
\usepackage{amsfonts}
\usepackage{amssymb}
\usepackage{graphicx}
\usepackage{mathrsfs}
\usepackage{bm}
\usepackage{color}
\usepackage{hyperref}
\usepackage[active]{srcltx}
\usepackage{cases}
\usepackage{ulem}
\usepackage{multirow}
\usepackage{braket}
\usepackage[utf8]{inputenc}

\begin{document}

\title{Absorption to fluctuating bunching states in non-unitary boson dynamics}

\author{Ken Mochizuki}
\affiliation{Department of Applied Physics, University of Tokyo, Tokyo 113-8656, Japan}
\affiliation{Nonequilibrium Quantum Statistical Mechanics RIKEN Hakubi Research Team, RIKEN Cluster for Pioneering Research (CPR), 2-1 Hirosawa, Wako 351-0198, Japan}

\author{Ryusuke Hamazaki}
\affiliation{Nonequilibrium Quantum Statistical Mechanics RIKEN Hakubi Research Team, RIKEN Cluster for Pioneering Research (CPR), 2-1 Hirosawa, Wako 351-0198, Japan}
\affiliation{RIKEN Interdisciplinary Theoretical and Mathematical Sciences Program (iTHEMS), 2-1 Hirosawa, Wako 351-0198, Japan}

\begin{abstract}
We show that noisy non-unitary dynamics of bosons drives arbitrary initial states into a novel \textit{fluctuating bunching state}, where all bosons occupy one time-dependent mode. 
We propose a concept of the noisy spectral gap, a generalization of the spectral gap in noiseless systems, and demonstrate that exponentially fast absorption to the fluctuating bunching state takes place asymptotically. 
The fluctuating bunching state is unique to noisy non-unitary dynamics with no counterpart in any unitary dynamics and non-unitary dynamics described by a time-independent generator. 
We also argue that the times of relaxation to the fluctuating bunching state obey a universal power law as functions of the noise parameter in generic noisy non-unitary dynamics. 
\end{abstract}

\maketitle

\section{Introduction}
\label{sec:introduction}
Long-time relaxation dynamics of a system is a central issue in non-equilibrium physics. 
Recent studies show that rich long-time behaviors arise in isolated, Floquet, and open systems. 
Spectral analysis has been an inevitable procedure to understand noiseless linear dynamics described by a time-independent generator. 
Indeed, various intriguing phenomena have been uncovered through the eigenvalues and eigenmodes of the generator. 
Examples of such phenomena include thermalization dynamics and its breakdown in isolated and Floquet systems in light of the eigenstate thermalization hypothesis \cite{berry1977regular,
deutsch1991quantum,
srednicki1994chaos,
rigol2008thermalization,
eckstein2009thermalization,
pal2010many-body,
steinigeweg2013eigenstate,
beugeling2014finite,
dalessio2014long-time,
kim2014testing,
lazarides2014equilibrium,
steinigeweg2014pushing,
nandkishore2015many,
d2016quantum,
moessner2017equilibration,
mori2018thermalization,
turner2018quantum,
yoshizawa2018numerical,
abanin2019colloquium,
sala2020ergodicity,
yang2020hilbert,
serbyn2021quantum,
sugimoto2021test,
moudgalya2022quantum,
yoshinaga2022emergence} and drastic changes of steady states and relaxation times through spectral transitions in open systems 
\cite{hatano1996localization,
makris2008beam,
guo2009observation,
ruter2010observation,
lin2011unidirectional,
miri2012large,
regensburger2012parity,
peng2014parity,
feng2014single,
hodaei2014parity,
zhu2014PT,
mochizuki2016explicit,
xiao2017observation,
gong2018discrete,
iemini2018boundary,
hamazaki2019non,
hanai2019non,
li2019observation,
longhi2019non,
takasu2020pt,
haga2023quasiparticles,
le2023volume,
mochizuki2023distinguishability}. 

In recent years, temporally noisy dynamics, where the generator includes time-dependent random or stochastic parameters, has gathered extensive attention and become experimentally feasible \cite{murch2013observing,roch2014observation,
arute2019quantum,
noel2022measurement,
koh2023measurement}. 
Such noisy systems offer a versatile platform to explore universal features and intriguing phenomena of quantum dynamics. 
Examples include information scrambling in unitary dynamics~\cite{nahum2017quantum,
khemani2018operator,
nahum2018operator,
rakovszky2018diffusive,
von2018operator,
zhou2019emergent,
bertini2020scrambling,
kuo2020markovian,
znidaric2022solvable} and transitions of entanglement entropy in non-unitary dynamics of monitored quantum systems \cite{li2018quantum,
cao2019entanglement,
chan2019unitary,
skinner2019measurement,
szyniszewski2019entanglement,
choi2020quantum,
fuji2020measurement,
gullans2020dynamical,
lunt2020measurement,
szyniszewski2020universality,
turkeshi2020measurement,
alberton2021entanglement,
lu2021spacetime,
noel2022measurement,
sierant2022universal,
granet2023volume,
koh2023measurement,
le2023volume,
yamamoto2023localization}. 
In stark contrast with noiseless cases, however, the direct spectral analysis for noisy dynamics has not been fully developed. 
This is because diagonalizing one generator does not bring full information for dynamics in noisy cases. 
While the spectral analysis for the averaged dynamics over noise is possible~\cite{kuo2020markovian,
nahum2018operator,
rakovszky2018diffusive,
von2018operator,
zhou2019emergent,
bertini2020scrambling,
znidaric2022solvable}, this procedure may discard some important information, such as how the state after long time is affected by the temporal noise.

\begin{figure}[t]
\begin{center}
\includegraphics[width=7cm]{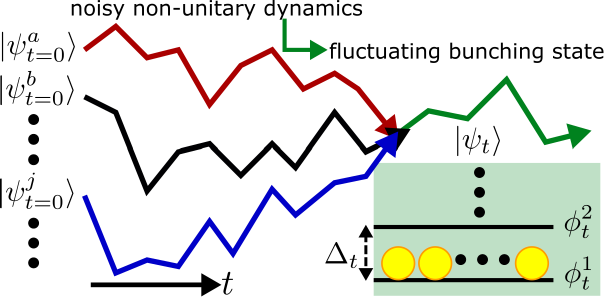}
\caption{ Schematic picture of the fluctuating bunching state. 
Various initial states, $\ket{\psi_{t=0}^a},\ket{\psi_{t=0}^b},\cdots$, are absorbed into the fluctuating bunching state, where all bosons described by yellow circles occupy one dominant eigenmode of $V_t$. 
We can define the time-dependent gap $\Delta_t$ from the eigenvalues of $\ln(V_t)/t$.}
\label{fig:fluctuating-bunching_state}
\end{center}
\end{figure}

In this work, we show that noisy non-unitary dynamics of free bosons results in a novel \textit{fluctuating bunching state}, where all bosons are absorbed to one time-dependent mode from any initial states after long times, as schematically shown in Fig. \ref{fig:fluctuating-bunching_state}. 
For this purpose, we propose a new analysis based on a \textit{noisy spectral gap}, a generalization of the spectral gap in noiseless systems. 
Using non-unitary free-boson dynamics subject to random losses and postselections, we demonstrate that the noisy spectral gap takes a non-zero value. 
This suggests the exponentially fast absorption to the fluctuating bunching state, which does not arise in any unitary dynamics and non-unitary dynamics by time-independent generators. 
We also argue that the relaxation times for the absorption are universally proportional to the inverse square of the noise strength. 

The rest of this paper is organized as follows. 
In Sec. \ref{sec:fluctuating-bunching_states}, we explain the general framework of the fluctuating bunching state, where the noisy spectral gap is also introduced. 
In Sec. \ref{sec:model}, we show that fluctuating bunching states emerge in a concrete model of an optical network with photon loss effect and postselection. 
In Sec. \ref{sec:relaxation-time}, we find the non-trivial power law of relaxation times and discuss that a wide range of non-unitary dynamics universally exhibit the power law. 
Section \ref{sec:conclusion} is devoted to the conclusion.

\section{Fluctuating bunching states in the general framework}
\label{sec:fluctuating-bunching_states}
We consider general non-unitary dynamics of free bosons where pure states are mapped to pure states and the number of bosons is conserved; if we set an initial state of $n$ bosons as $\ket{\psi_{t=0}}\propto\prod_{p=1}^n\hat{b}_{x_p^\mathrm{in}}^\dagger\ket{0}$, the quantum state after $t$-step discrete dynamics described by $V_t$ becomes \cite{mochizuki2023distinguishability}
\begin{align}
    \ket{\psi_t}=
    \frac{\prod_{p=1}^n
    \left(\sum_x[V_t]_{xx_p^\mathrm{in}}
    \hat{b}_x^\dagger\right)\ket{0}}
        {\sqrt{N_t}}.
    \label{eq:time-evolution}
\end{align}
Here, $p=1,2,\cdots,n$ is the label of each boson, $\hat{b}^\dagger_x$ is the creation operator of a boson at position $x\in[-X/2+1,X/2]$, $x_p^\mathrm{in}$ is the initial position of the $p$th boson, $\ket{0}$ is the vacuum state, and $\sqrt{N_t}$ is the normalization factor to ensure $\langle\psi_t|\psi_t\rangle=1$. 
We consider the non-unitary dynamical matrix $V_t$ decomposed as 
\begin{align}
    V_t=Q_tQ_{t-1} \cdots Q_2Q_1,
    \label{eq:V_t}
\end{align}
with $\{Q_t\}$ being non-unitary matrices for one step. 

We show that the spectrum of $V_t$ plays an important role in noisy dynamics with i.i.d.~random non-unitary matrices $\{Q_t\}$. 
To see this, assuming the diagonalizability, we consider the eigenequations of $V_t$, 
\begin{align}
    V_t\phi_t^i=\lambda_{t,i}\phi_t^i,\ \  
    V_t^\dagger\tilde{\phi}_t^i
    =\lambda_{t,i}^*\tilde{\phi}_t^i.
    \label{eq:eigen-value}
\end{align}
Here, $\phi_t^i\:(\tilde{\phi}_t^i)$ is the $i$th right (left) eigenmode for $V_t$ with the bi-orthonormality $\tilde{\phi}_t^{i\dagger}\phi_t^j=\delta_{ij}$, and eigenvalues $\{\lambda_{t,i}\}$ are ordered as $|\lambda_{t,1}|\geq|\lambda_{t,2}|\cdots\geq|\lambda_{t,X}|$. 

If $|\lambda_{t,1}|\gg|\lambda_{t,2}|$ is satisfied, $\ket{\psi_t}$ becomes
\begin{align}
    \ket{\psi_t}\simeq
    \frac{\left(\tilde{b}_{t,1}^\dagger\right)^n\ket{0}}
    {\sqrt{\tilde{N}_t}},
   \label{eq:long-time_limit}
\end{align}
where $\tilde{b}_{t,i}^\dagger=\sum_x\phi_t^i(x)\hat{b}_x^\dagger$ is a creation operator for an eigenmode $\phi_t^i$ with $\phi_t^{i\dagger}\phi_t^i=1$ and $\tilde{N}_t=\bra{0}\tilde{b}_{t,1}^n\tilde{b}_{t,1}^{\dagger n}\ket{0}$ is the normalization factor. 
We refer to Eq. (\ref{eq:long-time_limit}) as the bunching state because all bosons occupy a single mode $\phi_t^1$, as schematically shown in Fig. \ref{fig:fluctuating-bunching_state}. 
We note that, if the bunching state emerges, classical computers can easily sample the probability distribution of bosons by sampling that of distinguishable particles \cite{mochizuki2023distinguishability}. 
In the following, we show that bosons become distinguishable even in noisy non-unitary dynamics, while Ref. \cite{mochizuki2023distinguishability} explores noiseless dynamics. 
The classically computable distribution is contrasted to the unitary dynamics where the boson sampling problem can exhibit quantum supremacy \cite{aaronson2011computational,
aaronson2013the,
broome2013photonic,
crespi2013integrated,
spring2013boson,
tillmann2013experimental,
lund2014boson,
ticky2014stringent,
rahimi2015what,
rahimi2016sufficient,
hamilton2017gaussian,
he2017time,
loredo2017boson,
neville2017classical,
wang2017high,
zhong2020quantum}. 

Equation (\ref{eq:long-time_limit}) independent of $\{x_p^\mathrm{in}\}$ indicates the memory-loss effect, where all initial states are absorbed into the bunching state. 
Such memory loss of quantum states is absent for unitary dynamics, where $|\lambda_{t,i}|$ are the same for all eigenmodes and thus $\ket{\psi_t}$ always includes information about the initial state.

To understand the absorption into the bunching state in noisy non-unitary dynamics, we introduce the noisy spectral gap as 
\begin{align}
    \Delta=\lim_{t\rightarrow\infty}\overline{\Delta_t},\ 
    \Delta_t=-\ln\left(|\lambda_{t,2}/\lambda_{t,1}|\right)/t,
    \label{eq:gap}
\end{align}
where the overline denotes the average over random noises. 
Here, we assume that the sample fluctuation of $\Delta_t$ is small, $\left(\overline{\Delta_t^2}-\overline{\Delta_t}^2\right)^{1/2}\ll\overline{\Delta_t}$. 
Then, a non-zero value of $\Delta$ suggests the exponentially fast decay of $|\lambda_{t,2}/\lambda_{t,1}|$ for typical samples. 
Note that $\Delta$ is the generalization of the spectral gap $-\ln(|\lambda_{t,2}/\lambda_{t,1}|)/t=-\ln(|\kappa_2/\kappa_1|)$ in noiseless systems, where $\{\kappa_i\}$ with $|\kappa_i|\geq|\kappa_{i+1}|$ are eigenvalues of the one-step time-independent evolution matrix $Q_t=Q$. 
However, there are qualitative differences between noisy and noiseless systems.
One is that asymptotic convergence of $\overline{\Delta_t}$ is non-trivial in noisy dynamics while it is explicitly obtained in noiseless dynamics. 
This is because the spectrum of the time-dependent generator $Q_t$ does not give the gap in the noisy case, whereas that of the time-independent generator $Q$ does in the noiseless case.
We later demonstrate that $\Delta$ exists and takes non-zero values for a concrete noisy model. 
Another feature unique to noisy non-unitary dynamics, which is distinct from the noiseless dynamics through a time-independent generator, is that the system does not reach any stationary (i.e. time-independent) state but leads to the bunching state largely fluctuating in time. 
This is because the dominant eigenmode $\phi_t^1$ depends on time, unlike the noiseless case where $\{\phi_t^i\}$ corresponding to eigenmodes of $Q$ are independent of $t$.  
We note that such an absorption into the fluctuating state in quantum dynamics has not been known before.

We also explore relaxation times $\tau$ that bosonic systems take to reach the fluctuating bunching states. 
In noisy dynamics, where $Q_t$ involves a random variable $R$ satisfying $\overline{R}=0$, we argue that relaxation times typically exhibit a power law
\begin{align}
    \tau\propto\beta^{-2}
    \label{eq:relaxation-time_power}
\end{align}
for small $\beta$ with $\beta\propto(\overline{R^2})^{\frac{1}{2}}$ being the strength of the noise. 
This power law is discussed toward the end of the manuscript in detail. 
While we can evaluate relaxation times through various quantities, one natural choice is the definition from the noisy spectral gap, $\tau_\Delta\propto1/\Delta$. 

Before closing this section, we briefly mention the gap discrepancy problem. 
In the non-unitary dynamics by a time-independent generator, due to small overlaps of left and right eigenmodes the spectral gap obtained from the generator may not give plausible predictions of relaxation times \cite{mori2020resolving,
haga2021liouvillian,
znidaric2022solvable,
znidaric2023phantom}, which is referred to as the gap discrepancy problem. 
In the present work, where the spectral gap is generalized from noiseless dynamics to noisy dynamics, we consider cases where there is no gap discrepancy problem and thus the relaxation time can be evaluated through the noisy spectral gap.

\begin{figure*}[t]
\begin{center}
\includegraphics[width=17cm]{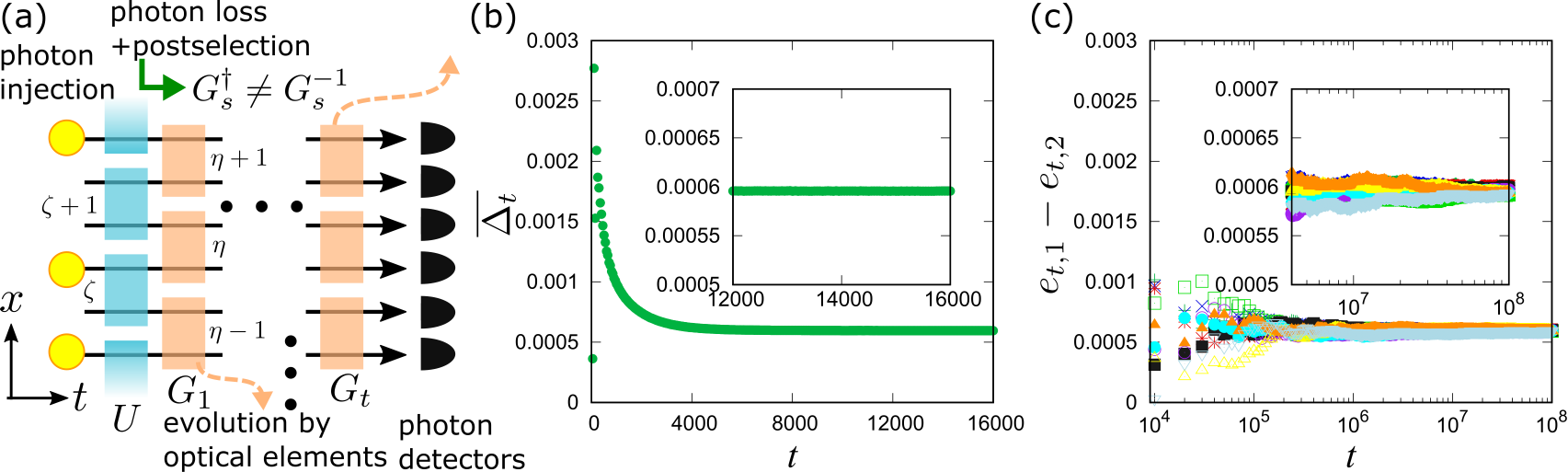}
\caption{(a) Schematic picture of the non-unitary dynamics of photons in an optical network. 
Photons evolve via the unitary matrix $U$ (blue rectangles) and then experience non-unitary dynamics by $G_t$ (orange rectangles) due to the photon loss effect and postselection. 
(b) The convergence of $\overline{\Delta_t}$ into the  noisy spectral gap $\Delta\simeq6\times10^{-4}$, where the number of samples is $10^5$. 
(c) The difference between the first and second Lyapunov exponents. 
Symbols with different colors correspond to different realizations of noises $\{z_{\eta,t}\}$. 
For large $t$, $e_{t,1}-e_{t,2}$ for each trajectory converges to the sample-independent value same as $\Delta$. 
In (b) and (c), $\beta=0.3$ and $X=20$.}
\label{fig:boson-sampling_exponent}
\end{center}
\end{figure*}

\section{Fluctuating bunching states in a concrete model}
\label{sec:model}
We demonstrate that the fluctuating bunching state emerges in noisy non-unitary dynamics whose time-evolution matrix for one step is given by 
\begin{align}
    Q_t=G_tU.
\end{align} 
Here, the time-independent unitary matrix $U$ is
\begin{align}
    U=\bigoplus_\zeta
    \left[\begin{array}{cc}
    e^{i\theta_1}\cos(\theta_3)&-e^{i\theta_2}\sin(\theta_3)\\
    e^{-i\theta_2}\sin(\theta_3)&e^{-i\theta_1}\cos(\theta_3)
    \end{array}\right],
    \label{eq:U}
\end{align}
where $\{\theta_j\}$ with $j=1,2,3$ are real parameters, $\zeta$ represents a two-site unit-cell covered by a blue rectangle in Fig. \ref{fig:boson-sampling_exponent} (a), and the periodic boundary condition is imposed. 
The time-dependent non-unitary matrix $G_t$ is
\begin{align}
    G_t=\bigoplus_\eta
    \left[\begin{array}{cc}
    \cosh\left(\beta z_{\eta,t}\right)&
    \sinh\left(\beta z_{\eta,t}\right)\\
    \sinh\left(\beta z_{\eta,t}\right)&
    \cosh\left(\beta z_{\eta,t}\right)
    \end{array}\right],
    \label{eq:G_t}
\end{align}
where $z_{\eta,t}$ takes random values uniformly sampled from the box distribution, $z_{\eta,t}\in[-1/2,+1/2]$. 
Here, $\eta$ represents a two-site unit-cell covered by an orange rectangle in Fig. \ref{fig:boson-sampling_exponent} (a). 
The real parameter $\beta$ determines the strength of noise leading to non-unitary dynamics, and $\beta z_{\eta,t}$ corresponds to $R$ in the previous section. 

Equation (\ref{eq:time-evolution}) describes the dynamics of photons in an optical network with photon loss effect and postselection. 
The creation operator $\hat{b}_x^\dagger$ is transformed by $U$ when photons pass through linear optical elements, such as beam splitters, phase shifters, and wave plates. 
Non-unitary dynamics by $G_t$ occurs when we postselect the cases where all photons remain in the system, despite the effect of photon loss. 
The photon loss can be artificially introduced through optical elements coupled to the environment \cite{mochizuki2023distinguishability}. 
We consider situations where the optical elements have imperfections leading to noise, that is, the strength of loss randomly deviates from a mean value depending on $\eta$ and $t$. 
This type of noise makes our setting distinct from conventional regimes such as fluctuational electrodynamics treating thermal noises and quantum fluctuations \cite{volokitin2007near-field}. 
We note that the dynamics with $n$ conserved is distinct from the dynamics averaged over all outcomes where the number of photons can be decreased, like the dynamics governed by the Gorini-Kossakowski-Sudarshan-Lindblad equation \cite{breuer2002theory}.
The long-time limit of the averaged dynamics becomes the trivial vacuum state with $n=0$, which is contrasted to our setting where the fluctuating bunching states emerge.

Figure \ref{fig:boson-sampling_exponent} (b) shows that $\overline{\Delta_t}$ converges to a constant value $\Delta$ in time, where the overline denotes the average over random realizations of $\{z_{\eta,t}\}$. 
The non-zero gap $\Delta$ implies that exponentially fast absorption to the fluctuating bunching state asymptotically occurs for this model. 
We note that $\Delta$ is proportional to $1/X$, which suggests that the fluctuating bunching state emerges with the timescale $\tau_\Delta=\mathcal{O}(X)$, as detailed in Appendix \ref{sec:size-dependence}. 
We note that the non-zero noisy spectral gap in this paper is defined for each finite-size system, and we do not consider the thermodynamic limit $X\rightarrow\infty$.

\begin{figure}
\begin{center}
\includegraphics[width=6.5cm]{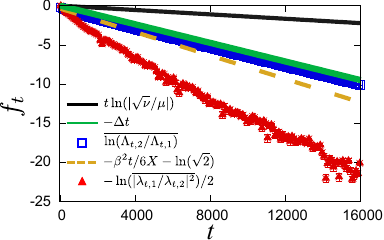}
\caption{Decays of $f_t=-\ln\left(\overline{|\lambda_{t,1}/\lambda_{t,2}|^2}\right)/2$ (red symbols), $\overline{\ln\left(\Lambda_{t,2}/\Lambda_{t,1}\right)}$ (blue symbols), and $- \Delta t$ (the green line), respectively, which indicate that the fluctuating bunching state emerges after long-time dynamics. 
The average is taken over $10^5$ samples with $\beta=0.3$ and $X=20$. 
The orange broken line is $f_t=-\beta^2t/6X-\ln(\sqrt{2})$ obtained from Eq. (\ref{eq:log-omega_average_expansion}), which is calculated through the perturbation analysis on $\overline{\ln\left(\Lambda_{t,2}/\Lambda_{t,1}\right)}$. 
The black line $f_t=t\ln(\sqrt{\nu}/\mu)$ represents the analytical upper bound on $-\ln\left(\overline{|\lambda_{t,1}/\lambda_{t,2}|^2}\right)/2$, derived from Eq. (\ref{eq:inequality_mu-nu}). 
}
\label{fig:ratio}
\end{center}
\end{figure}

We find evidence that $\Delta$ gives the decay rate to the fluctuating bunching state even for each trajectory. 
To see this, we evaluate the long-time behavior of $e_{t,1}-e_{t,2}$ as an approximation of $\Delta$ at $t=10^8$. 
Here, $e_{t,1}$ and $e_{t,2}$ are the first and second instantaneous Lyapunov exponents defined as
\begin{align}
    e_{t,i}=\ln(\Lambda_{t,i})/t
    \label{eq:Lyapunov-exponent}
\end{align}
without ensemble average, where $\{\Lambda_{t,i}\}$ are singular values of $V_t$,
\begin{align}
    \tilde{V}_t\Phi_t^i=\Lambda_{t,i}^2\Phi_t^i
    \label{eq:singular-values}
\end{align}
with $\tilde{V}_t=V^\dagger_tV_t$ \cite{crisanti2012products}. 
Singular values $\{\Lambda_{t,i}\}$ are ordered as $\Lambda_{t,1} \geq \Lambda_{t,2} \geq \cdots \geq \Lambda_{t,X}$. 
We can also capture relaxation dynamics through $\{\Lambda_{t,i}\}$ as well as $\{\lambda_{t,i}\}$. 
Indeed, both $|\lambda_{t,1}|\gg|\lambda_{t,2}|$ and  $\Lambda_{t,1}\gg\Lambda_{t,2}$ indicate that $V_t$ can be appoximated by the rank-$1$ matrix. 
As shown in Fig. \ref{fig:ratio}, $\Lambda_{t,2}/\Lambda_{t,1}$ exponentially decays when $t$ is increased. 
Since limitations on the numerical precision and the exponential decay of $\Lambda_{t,2}/\Lambda_{t,1}$ prevent us from directly computing $\Lambda_{t,2}/\Lambda_{t,1}$ for large $t$, we obtain $e_{t,1}-e_{t,2}$ through evaluating growth rates of randomly chosen vectors, as detailed in Appendix \ref{sec:numerics_Lyapunov-exponents}. 
Figure~\ref{fig:boson-sampling_exponent} (c) shows that $e_{t,1}-e_{t,2}$ converges to the same value as $\Delta$ for all samples, which indicates $|\lambda_{t,2}/\lambda_{t,1}| \sim e^{- \Delta t}$ for each random realization. 
Note that the spectral gap and Lyapunov exponents of noisy non-unitary quantum dynamics have seldom been explored, while the first Lyapunov exponents for products of transfer matrices have been studied for obtaining localization lengths in time-independent systems with random potentials \cite{kramer1993localization}.

\begin{figure}
\begin{center}
\includegraphics[width=6.5cm]{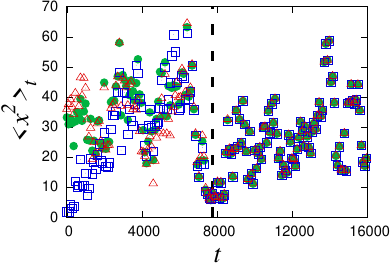}
\caption{Expectation values of $\sum_xx^2\hat{b}^\dagger_x\hat{b}_x/n$ under non-unitary dynamics of three bosons with different initial states. 
Green, blue, and red symbols respectively correspond to   $(x_1^\mathrm{in},x_2^\mathrm{in},x_3^\mathrm{in})=(-6,1,8)$, $(-1,0,1)$, and $(-6,-5,6)$ with a fixed sample $\{z_{\eta,t}\}$. 
The three trajectories are absorbed into one time-dependent trajectory independent of initial states in the long-time regime, while they depend on initial states in the short-time regime. 
The black broken line represents $t=\tau_\Delta=|\ln(c)|/\Delta$ with $c=10^{-2}$ at which $\overline{\ln(|\lambda_{t,2}/\lambda_{t,1}|)} =\ln(c)$ is satisfied. 
The parameters are $(\theta_1,\theta_2,\theta_3)=(0.37\pi,0.19\pi,0.25\pi)$ and $\beta=0.3$, with $X=20$.}
\label{fig:trajectory_x}
\end{center}
\end{figure}

As a consequence of the absorption to the fluctuating bunching state, physical observables exhibit fluctuating dynamics independent of initial states in the long run. 
Figure~\ref{fig:trajectory_x} shows that expectation values of an observable, $\left<x^2\right>_t=\bra{\psi_t}\sum_xx^2\hat{b}^\dagger_x\hat{b}_x\ket{\psi_t}/n$, approach the same fluctuating trajectory with large $t$ for all initial states. 
We note that the dominant mode $\phi_t^1$ is spatially localized as detailed in Appendix \ref{sec:localization}, where the localization position depends on time as indicated by Fig. \ref{fig:trajectory_x}. 
The fluctuation and localization of the bunching state mean that there is no equilibration and thermalization.

\section{Power law of relaxation times}
\label{sec:relaxation-time}
We argue that relaxation times at which the bosonic system reaches the fluctuating bunching state are universally given by $\tau\propto\beta^{-2}$ for small noise strength $\beta$. 
Our argument follows from two steps. 
First, we introduce four versions of relaxation times, $\tau_\Delta$, $\tau_\lambda$, $\tau_\Lambda$, and $\tau_x$, and numerically show that they exhibit similar behavior, in particular, the power law in Eq. (\ref{eq:relaxation-time_power}). 
Second, we analytically evaluate $\tau_\lambda$ and $\tau_\Lambda$ instead of $\tau_\Delta$, for which the analytical estimation is difficult. 
In addition, we make plausible arguments for the universality of the power law $\tau\propto\beta^{-2}$.

\begin{figure}
\begin{center}
\includegraphics[width=6.5cm]{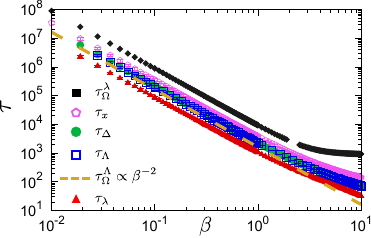}
\caption{Relaxation times as functions of the noise parameter, which exhibit the power law in Eq. (\ref{eq:relaxation-time_power}). 
Green, blue, red, and purple symbols represent $\tau_\Delta$, $\tau_\Lambda$, $\tau_\lambda$, and $\tau_x$, respectively, with $c=10^{-6}$ and $X=20$. 
Here, the average in $f_t$ is taken over $10^2$ samples for obtaining one $\tau$, and then we iterate it $10^2$ times and obtain the averaged $\tau$ over them. 
Expectation values are computed with $n=2$, where initial states are $(x_1^\mathrm{in},x_2^\mathrm{in})=(-5,5)$ for $\left<x^2\right>_{t,a}$ and $(-1,0)$ for $\left<x^2\right>_{t,b}$. 
The black symbols and orange broken line correspond to $\tau_\Omega^\lambda$ and $\tau_\Omega^\Lambda$, in Eqs. (\ref{eq:relaxation-time_CS}) and (\ref{eq:relaxation-time_perturbation}), respectively, where the latter is exactly proportional to $\beta^{-2}$. }
\label{fig:relaxation-time}
\end{center}
\end{figure}

As the first step, we define relaxation times as the smallest time steps at which $f_t\leq\ln(c)$ is satisfied, where $c\ll1$ is a threshold and a function $f_t$ is defined through a quantity related to $|\lambda_{t,2}/\lambda_{t,1}|,\ \Lambda_{t,2}/\Lambda_{t,1}$, or observables. 
On the basis of the noisy spectral gap, we define the relaxation time $\tau_\Delta$ with $f_t=-\Delta t$, which leads to $\tau_\Delta=|\ln(c)/\Delta|$. 
Figure \ref{fig:ratio} shows that other two quantities, $-\ln[\overline{|\lambda_{t,1}/\lambda_{t,2}|^2}]/2$ and $\overline{\ln\left(\Lambda_{t,2}/\Lambda_{t,1}\right)}$, also decay in time. 
In particular, behaviors of $- \Delta t$ and $\overline{\ln(\Lambda_{t,2}/\Lambda_{t,2})}$ are quite similar. 
Then, we define relaxation times $\tau_\lambda$ and $\tau_\Lambda$ through $f_t=-\ln\left(\overline{|\lambda_{t,1}/\lambda_{t,2}|^2}\right)/2$ and  $f_t=\overline{\ln(\Lambda_{t,2}/\Lambda_{t,1})}$, respectively. 
In Fig.~\ref{fig:relaxation-time}, we discover that $\tau_\lambda$, $\tau_\Lambda$, and $\tau_\Delta$ all show the same $\beta$ dependence, $\tau\propto\beta^{-2}$, for sufficiently small $\beta$. 
In particular, $\tau_\Lambda$ and $\tau_\Delta$ are almost overlapped. 
We also find that the same power law appears for an additional timescale $\tau_x$, which is obtained from $f_t=\ln\left(\overline{|\left<x^2\right>_{t,a}-\left<x^2\right>_{t,b}|}\right)$. 
Here, the subscripts for $\left<x^2\right>_t$ represent the different initial states.

As the second step, we semi-analytically give an upper bound of $\tau_\lambda$.
To this end, we focus on $\Omega_t^\lambda=|\mathrm{tr}(V_t)^2-\mathrm{tr}(V_t^2)|^2/|\mathrm{tr}(V_t)|^4$, which becomes $\Omega_t^\lambda\simeq4|\lambda_{t,2}/\lambda_{t,1}|^2$ if eigenvalues satisfy $|\lambda_{t,1}|\gg|\lambda_{t,2}|\gg|\lambda_{t,3}|$. 
Thus, $\tau_\lambda$ can be evaluated through $\ln(1/\Omega_t^\lambda)\geq\ln(1/4c^2)$. 
As detailed in Appendix \ref{sec:Cauchy-Schwarz_inequality}, we find a lower bound 
\begin{align}
|\mu^2/\nu|^t\leq\overline{4/\Omega_t^\lambda},
\label{eq:inequality_mu-nu}
\end{align}
which is satisfied asymptotically. 
Here, $\mu$ and $\nu$ are respectively the largest eigenvalues of $\mathcal{Q}=\overline{Q_t \otimes Q_t^*}$ and $\mathbb{Q}\mathbb{S}/4=\overline{(Q_t \otimes Q_t)\otimes(Q_t \otimes Q_t)^*}\mathbb{S}/4$, with $\mathbb{S}$ being a complicated matrix independent of $\beta$. 
Thus, we define the relaxation time $\tau_\Omega^\lambda$ through $f_t=\ln\left(|\sqrt{\nu}/\mu|^t\right)$, which leads to $\tau_\Omega^\lambda=|\ln(c)/\ln(|\sqrt{\nu}/\mu|)|$. 
In Figs. \ref{fig:ratio} and \ref{fig:relaxation-time}, $\ln\left(|\sqrt{\nu}/\mu|^t\right)$ and $\tau_\Omega^\lambda$ (black) are respectively above $-\ln(\overline{|\lambda_{t,1}/\lambda_{t,2}|^2})/2$ and $\tau_\lambda$ (red) due to the inequality (\ref{eq:inequality_mu-nu}).  
Figure~\ref{fig:relaxation-time} indicates the power law of $\tau_\Omega^\lambda$,
\begin{align}
    \tau_\lambda\leq\tau_{\Omega}^\lambda\propto\beta^{-2}.
    \label{eq:relaxation-time_CS}
\end{align}
This is because $\mathcal{Q}$ ($\mathbb{Q}$) can be expanded as $\mathcal{Q}\simeq\mathcal{U}+\beta^2\mathcal{F}$ ($\mathbb{Q}\simeq\mathbb{U}+\beta^2\mathbb{F}$) for small $\beta$ with $\mathcal{U}$ and $\mathcal{F}$ ($\mathbb{U}$ and $\mathbb{F}$) being some complicated but $\beta$-independent matrices whose explicit forms are written in Appendix \ref{sec:Cauchy-Schwarz_inequality}. 
Since $\ln(|\sqrt{\nu}/\mu|)$ should be zero for $\beta=0$, we believe that the subleading terms $\beta^2$ of $\mathcal{Q}$ and $\mathbb{Q}$ typically lead to $\ln(|\sqrt{\nu}/\mu|)\propto\beta^2$. 
While $\tau_\Omega^\lambda$ only gives the upper bound of $\tau_\lambda$,  Fig. \ref{fig:relaxation-time} shows that both timescales are proportional to $\beta^{-2}$ in this model. 

We also analytically obtain an approximated value of $\tau_\Lambda$ using a perturbative analysis. 
To this end, we focus on $\Omega_t^\Lambda=[\mathrm{tr}(\tilde{V}_t)^2-\mathrm{tr}(\tilde{V}_t^2)]/\mathrm{tr}(\tilde{V}_t)^2$ with $\tilde{V}_t=V^\dagger_tV_t$, which leads to $\Omega_t^\Lambda\simeq2(\Lambda_{t,2}/\Lambda_{t,1})^2$ if $\Lambda_{t,1}\gg\Lambda_{t,2}\gg\Lambda_{t,3}$ is satisfied. 
Thus, $\tau_\Lambda$ can be evaluated through $\ln(\Omega_t^\Lambda)\leq\ln(2c^2)$. 
When $\beta$ is small, we can neglect higher-order terms of $\beta$, which results in $G_t\simeq\bigoplus_\eta(\sigma_0+\beta z_{\eta,t}\sigma_1+\beta^2z_{\eta,t}^2\sigma_0/2)$, where $\sigma_0$ and $\sigma_1$ are the Pauli matrices. 
Thus, we can approximate $Q_t$ as
\begin{align}
    Q_t \simeq A_t + \beta B_t+ \beta^2 C_t,
    \label{eq:Qt_expansion}
\end{align}
where $A_t=U,\ B_t=U\bigoplus_\eta z_{\eta,t}\sigma_1$, and $C_t=U\bigoplus_\eta z_{\eta,t}^2\sigma_0/2$. 
As detailed in Appendix \ref{sec:perturbation}, this expansion leads to 
\begin{align}
   \overline{\ln\left(\Omega_t^\Lambda\right)}
   &\simeq-\mathrm{tr}\left[2\overline{B^\dagger_s B_s}
   +\overline{(A^\dagger_s B_s)^2}
   +\overline{(B^\dagger_s A_s)^2}\right]
   \frac{\beta^2}{X^2}t\nonumber\\
   &=-\frac{\beta^2}{3X}t
    \label{eq:log-omega_average_expansion}
\end{align}
which is valid when $\beta^2t/X$ is small. 
The orange broken line in Fig. \ref{fig:relaxation-time} corresponds to
\begin{align}
    \tau_\Omega^\Lambda=6X|\ln(c)+\ln(\sqrt{2})|\beta^{-2},
    \label{eq:relaxation-time_perturbation}
\end{align}
defined by $f_t=-\beta^2/6X-\ln(\sqrt{2})\simeq\overline{\ln(\Lambda_{t,2}/\Lambda_{t,1})}$. 
$\tau_\Omega^\Lambda$ is parallel to $\tau_\Lambda$, $\tau_\lambda$, and $\tau_x$, which is consistent with our argument that all relaxation times satisfy $\tau\propto\beta^{-2}$. 

We argue that the power law in Eq. (\ref{eq:relaxation-time_power}) for small $\beta$ emerges in a wide range of noisy non-unitary dynamics, on the basis of the following two assumptions satisfied in our model. 
We first assume that $\ln(|\sqrt{\nu}/\mu|)\propto\beta^2$ and thus $\tau_\Omega^\lambda\propto\beta^{-2}$, which is plausible because subleading order terms of $\mathcal{Q}$ and $\mathbb{Q}$ become $\beta^2$ when the average of noise is zero, as detailed in Appendix \ref{sec:perturbation}. 
We also assume that various relaxation times such as $\tau_\lambda$, $\tau_\Omega^\lambda$, $\tau_\Delta$, and $\tau_\Omega^\Lambda$ exhibit the same dependence on the noise parameter $\beta$, as confirmed in the concrete model.  
In addition, details of models have no effect on the power law $\tau_\Omega^\Lambda\propto\beta^{-2}$, as long as the average of the noise is zero.  
We indeed discover the same power law $\tau\propto\beta^{-2}$ for a model proposed in Ref. \cite{chen2021non}, which is different from that in the present work, as shown in Appendix \ref{sec:chen-model}.

\section{Conclusion}
\label{sec:conclusion}
We have shown that the novel fluctuating bunching state, where all bosons occupy one time-dependent mode independent of initial states, appears in noisy non-unitary dynamics. 
This is analyzed through the spectrum of noisy non-unitary time-evolution matrices.  
In particular, the noisy spectral gap $\Delta$, the extension of the spectral gap in noiseless systems, plays a key role.  
We have also argued that times of relaxation to the fluctuating bunching state universally exhibit the power law as functions of the noise parameter, $\tau\propto\beta^{-2}$. 

The spectral analysis can be applied to general noisy dynamics, not restricted to the dynamics of bosons explored in this work. 
For example, our analysis should be applicable to the dynamics of fermions, while fermions do not occupy one dominant mode due to the Pauli exclusion rule. 
In addition, it may be possible to analyze transitions of relaxation times, such as purification transitions in monitored quantum many-body systems \cite{gullans2020dynamical}, through the noisy spectral gap $\Delta$ in the present work.

\begin{acknowledgements}
We thank Hideaki Obuse, Takashi Mori, and Kazuki Sone for fruitful discussions. 
We also appreciate Zongping Gong, Yohei Fuji, and Hisanori Oshima for valuable comments on the manuscript. 
This work is supported by JST ERATO Grant Number JPMJER2302. 
Ken Mochizuki is supported by JSPS KAKENHI Grant Number JP23K13037.
\end{acknowledgements}

\bibliographystyle{apsrev4-2}
\bibliography{reference.bib}

\appendix
\onecolumngrid

\section{System-size dependence of the noisy spectral gap}
\label{sec:size-dependence}
\begin{figure}
\begin{center}
\includegraphics[width=7cm]{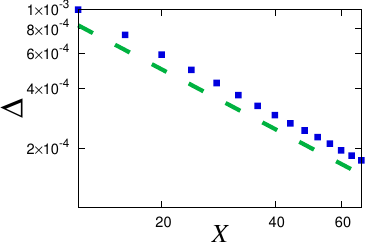}
\caption{ The noisy spectral gap as a function of $X$. 
The blue symbols are numerically obtained $\Delta$ with $(\theta_1,\theta_2,\theta_3)=(0.37\pi,0.19\pi,0.25\pi)$ and $\beta=0.3$. 
The green broken line is $10^{-2}/X$.}
\label{fig:gap_size-dependence}
\end{center}
\end{figure}

The noisy spectral gap $\Delta$ becomes small as we increase the system size $X$ for a system studied in the main text. 
Figure \ref{fig:gap_size-dependence} shows that $\Delta$ is proportional to $1/X$ for this model. 
This means that bosonic states are absorbed into the fluctuating bunching state with the timescale $\tau_\Delta \sim X$, and the system is gapless in the thermodynamic limit $X\rightarrow\infty$.

\section{Numerical calculation of Lyapunov exponenets}
\label{sec:numerics_Lyapunov-exponents}
The exponential decay of $\Lambda_{t,2}/\Lambda_{t,1}$ prevents us from directly computing the Lyapunov exponents through the diagonalization of $\tilde{V}_t$ for large $t$, due to limitations on the precision of numerics. 
Instead, we can compute the Lyapunov exponents by focusing on the dynamics of vectors, which is presented in Ref. \cite{crisanti2012products}. 
First, we choose two vectors $w_0^1$ and $w_0^2$ randomly. 
After that, these vectors are evolved as
\begin{align}
    w_{s+1}^1&=u_{s+1}/|u_{s+1}|,\ 
    u_{s+1}=\tilde{Q}_{t,M}w_s^1,
    \label{eq:dynamics_first}\\
    w_{s+1}^2&=v_{s+1}/|v_{s+1}|,\ 
    v_{s+1}=(I_X-\Pi_{s+1})\tilde{Q}_{t,M}w_s^2,
    \label{eq:dynamics_second}
\end{align}
where $\tilde{Q}_{t,M}=Q_{t+M}Q_{t+M-1} \cdots Q_{t+1},\ t=sM$, and $I_X$ is the $X \times X$ identity matrix. 
Here, $\Pi_s=[w_s^1][w_s^1]^\dagger$ is the projection matrix onto the direction of $w_s^1$, and thus $w_s^2$ is orthogonal to $w_s^1$. 
The Lyapunov exponents are obtained through
\begin{align} 
    e_1&=\lim_{T\rightarrow\infty}e_{t,1}
    =\lim_{T\rightarrow\infty}
    \frac{1}{TM}\sum_{s=1}^T\ln[\mathrm{VOL}_1(u_s)],
    \label{eq:lyapunov-exponent_first}\\
    e_1+e_2&=\lim_{T\rightarrow\infty}(e_{t,1}+e_{t,2})
    =\lim_{T\rightarrow\infty}\frac{1}{TM}
    \sum_{s=1}^T\ln[\mathrm{VOL}_2(u_s,v_s)],
    \label{eq:lyapunov-exponent_second}
\end{align}
where $t=TM$, $\mathrm{VOL}_1(u_s)=|u_s|$, and $\mathrm{VOL}_2(u_s,v_s)=|u_s||v_s|\sin(\theta_s)$ with $\cos(\theta_s)=|u^\dagger_sv_s|/|u_s||v_s|$. 
The right-hand sides of Eqs. (\ref{eq:lyapunov-exponent_first}) and (\ref{eq:lyapunov-exponent_second}) quantify the growth rates of two randomly chosen vectors, and thus correspond to the largest and second-largest Lyapunov exponents. 
This is because increasing $s$ makes $w_s^1$ ($w_s^2$) approach the first (second) Lyapunov vector, the eigenmode of $\lim_{t\rightarrow\infty}\tilde{V}_t$ corresponding to the largest (second-largest) eigenvalue. 
Since the vectors $w_s^1$ and $w_s^2$ are normalized for every $M$ step, we can make $t$ much larger than a time step at which $\Lambda_{t,2}/\Lambda_{t,1}$ becomes smaller than the numerical precision. 
Indeed, numerical results with $M=10^4$ and $T=10^4$ in Fig. \ref{fig:boson-sampling_exponent} (c) give a plausible value of the noisy spectral gap for $t=10^8$, while the directly computed $\Lambda_{t,2}/\Lambda_{t,1}$ becomes smaller than our numerical accuracy $10^{-16}$ in a range $10^4 \leq t \leq 10^5$, as indicated by Fig. \ref{fig:ratio}. 

\section{Localization of the bunching state}
\label{sec:localization}
\begin{figure}
\begin{center}
\includegraphics[width=12cm]{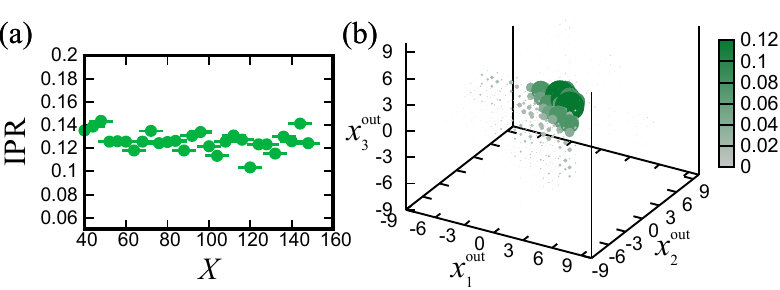}
\caption{(a) The averaged inverse participation ratio (IPR) of the dominant mode $\overline{\sum_x|\phi_t^1(x)|^4/[\sum_x|\phi_t^1(x)|^2]^2}$ as a function of the system size $X$. 
The IPR is almost independent of $X$, which indicates that $\phi_t^1$ is localized. 
The average is taken over $40$ samples and $100$ time steps after a time at which  $\overline{|\lambda_{t,2}/\lambda_{t,1}|}<c=10^{-6}$ is satisfied. 
(b) The probability distribution $P_t(\{x_q^\mathrm{out}\})$ of a bunching state with $n=3$, $(x_1^\mathrm{in},x_2^\mathrm{in},x_3^\mathrm{in})=(-6,1,8)$, $\beta=0.3$, and $t=42418$. 
The value of $P_t(\{x_q^\mathrm{out}\})$ is represented as the size and color of the symbols: the big (tiny) and dense green (shallow gray) symbols correspond to large (small) $P_t(\{x_q^\mathrm{out}\})$.}
\label{fig:IPR-distribution}
\end{center}
\end{figure}

We find that the occupied state $\phi_t^1$ is spatially localized in the model explored in the main text; the average of the inverse participation ratio (IPR) $\overline{\sum_x|\phi_t^1(x)|^4/[\sum_x|\phi_t^1(x)|^2]^2}$ does not decrease monotonically when we increase the system size $X$, as shown in Fig. \ref{fig:IPR-distribution} (a).  
Figure \ref{fig:IPR-distribution} (b) shows the probability distribution \cite{mochizuki2023distinguishability}
\begin{align}
    P_t(\{x^\mathrm{in}_p\},\{x^\mathrm{out}_q\})
    =\frac{|\mathrm{Per}[W_t]|^2}
   {N_t\prod_xn^\mathrm{in}_x!n^\mathrm{out}_x!},
   \label{eq:probability-distribution}
\end{align}
with $n=3$ and $t=42418$, at which  $|\lambda_{t,1}|\gg|\lambda_{t,2}|$. 
In addition, $n_x^\mathrm{in(out)}$ with $\sum_xn_x^\mathrm{in(out)}=n$ is the number of bosons at $x$ regarding the input (output) configuration of bosons, and $\mathrm{Per}[W_t]$ is the permanent of an $n$-by-$n$ matrix $W_t$ defined as the submatrix of $V_t$, $[W_t]_{qp}=[V_t]_{x_q^\mathrm{out}x_p^\mathrm{in}}$. 
Here, $x_q^\mathrm{in(out)}$ with $q=1,2,\cdots,n$ is the position of the $q$th input (output) boson. 
If the fluctuating bunching state is realized, initial states have no effect on the probability distribution, and thus $P_t(\{x^\mathrm{out}_q\})$ is independent of $\{x^\mathrm{in}_p\}$. 
We can see that the three bosons gather around one position due to the localization of $\phi_t^1$. 
Note that the localization position depends on $t$, as indicated by Fig. \ref{fig:trajectory_x} in the main text. 

\section{Power law of various relaxation times}
\label{sec:relaxation-times}
As discussed in the main text, we conjecture that various types of possible relaxation times for the absorption exhibit the same power-law dependence on $\beta$. 
In this section, we numerically confirm this conjecture for the model introduced in the main text.

\begin{figure}[htbp]
\begin{center}
\includegraphics[width=7cm]{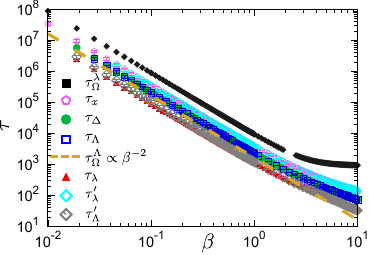}
\caption{Various relaxation times as functions of the noise parameter $\beta$, all of which exhibit the power law $\tau\propto\beta^{-2}$. 
Blue, red, purple, green, cyan, and grey symbols respectively represent relaxation times defined through $f_\tau\leq\ln(c)$ with $c=10^{-6}$, where $f_t=\overline{\ln\left(\Lambda_{t,2}/\Lambda_{t,1}\right)}$, $f_t=-\ln\left(\overline{|\lambda_{t,1}/\lambda_{t,2}|^2}\right)/2$, $f_t=\ln\left(\overline{|\left<x^2\right>_{t,a}-\left<x^2\right>_{t,b}|}\right)$ with $n=2$, $f_t=- \Delta t$,  $f_t=\ln\left(\overline{|\lambda_{t,2}/\lambda_{t,1}|^2}\right)/2$, and $f_t=-\ln\left(\overline{|\Lambda_{t,1}/\Lambda_{t,2}|^2}\right)/2$. 
The average included in $f_t$ is taken over $10^2$ samples for obtaining one $\tau$.
Then, we reiterate obtaining $\tau$ for $10^2$ times and take the average over them. 
The subscripts $a$ and $b$ for $\left<x^2\right>_t$ correspond to two initial states of bosons, $(x_1^\mathrm{in},x_2^\mathrm{in})=(-5,5)$ and $(-1,0)$.  
Regarding $\tau_\Delta$ with $f_t=-\Delta t$, $\Delta$ is obtained through the ensemble average over $10^4$ samples. 
The orange broken line corresponds to the relaxation time $\tau_\Omega^\Lambda$ defined through the perturbative analysis on $\overline{\ln\left(\Lambda_{t,2}/\Lambda_{t,1}\right)}$.
Since this relaxation time is obtained as $-\beta^2\tau_\Omega^\Lambda/6X-\ln(\sqrt{2})=\ln(c)$, the orange broken line is exactly proportional to $\beta^{-2}$. 
The black symbols represent $\tau_\Omega^\lambda=\ln(c)/\ln\left(|\sqrt{\nu}/\mu|\right)$, which is above the red symbols for large $t$, due to the inequality obtained in Eq. (\ref{eq:inequality_omega}). 
The parameters are $(\theta_1,\theta_2,\theta_3)=(0.37\pi,0.19\pi,0.25\pi)$ with the system size being $X=20$. }
\label{fig:relaxation-time_supple}
\end{center}
\end{figure}

We define a relaxation time at which the fluctuating bunching state is realized as the smallest time at which $f_\tau\leq\ln(c)$ is satisfied with $c\ll1$. 
In the main text, we explore four relaxation times $\tau_\lambda$, $\tau_\Lambda,$ $\tau_\Delta$, and $\tau_x$ respectively with $f_t=-\ln\left(\overline{|\lambda_{t,1}/\lambda_{t,2}|^2}\right)/2$,  $f_t=\overline{\ln\left(\Lambda_{t,2}/\Lambda_{t,1}\right)}$, $f_t=-\Delta t$, and $f_t=\ln\left(\overline{|\left<x^2\right>_{t,a}-\left<x^2\right>_{t,b}|}\right)$, where the number of bosons is $n=2$ and subscripts $a$ and $b$ represent different initial states. 
In Fig. \ref{fig:relaxation-time_supple}, these relaxation times correspond to red, blue, green, and purple symbols, respectively. 
In addition to these relaxation times explored in the main text, Fig. \ref{fig:relaxation-time_supple} also shows relaxation times $\tau_\lambda'$ with $f_t=\ln\left(\overline{|\lambda_{t,2}/\lambda_{t,1}|^2}\right)/2$ and $\tau_\Lambda'$ with $f_t=-\ln\left(\overline{|\Lambda_{t,1}/\Lambda_{t,2}|^2}\right)/2$, which respectively correspond to cyan and grey symbols.  
From Fig. \ref{fig:relaxation-time_supple}, we can understand that all relaxation times exhibit the same power law $\tau\propto\beta^{-2}$; 
 the orange broken line $\tau_\Omega^\Lambda=3X|2\ln(c)+\ln(2)|\beta^{-2}$, derived in Sec. \ref{sec:perturbation}, is exactly proportional to $\beta^{-2}$, and all
relaxation times are parallel to this line. 
From Fig. \ref{fig:relaxation-time_supple}, we can also understand that behaviors of the eigenvalues and singular values are quite similar; relaxation times based on $- \Delta t$ (green symbols) and $\overline{\ln\left(\Lambda_{t,2}/\Lambda_{t,1}\right)}$ (blue symbols) are almost identical. 

\section{analysis through the upper bound on $\tau_\lambda$}
\label{sec:Cauchy-Schwarz_inequality}
We evaluate the relaxation time $\tau_\lambda$ based on eigenvalues of $V_t$,
\begin{align}
    V_t\phi_t^i=\lambda_{t,i}\phi_t^i,
    \label{eq:eigen-value}
\end{align}
where $|\lambda_{t,1}|\geq|\lambda_{t,2}|\cdots\geq|\lambda_{t,X}|$. 
If $|\lambda_{t,2}/\lambda_{t,1}|$ is smaller than the threshold value $c\ll1$, the bunching state is realized within the precision $c$. 

As discussed in the main text, the timescale $\tau_\lambda$, which is obtained from $f_t=-\ln\left(\overline{|\lambda_{t,1}/\lambda_{t,2}|^2}\right)/2$, has an analytical upper bound $\tau_\Omega^\lambda$, which typically shows $\beta^{-2}$ decay.
In this section, we show this fact in detail.

\subsection{General bound}
\label{subsec:CS_general}
In this subsection, we give discussions applicable to a wide range of models that exhibit noisy non-unitary dynamics, not restricted to the model explored in the main text. 
Then, in the next subsection, we demonstrate the concrete application of our general  discussions for the model in the main text. 

To discuss the decay of $|\lambda_{t,2}/\lambda_{t,1}|$, we focus on $\Omega_t^\lambda$ defined as
\begin{align}
    \Omega_t^\lambda=\frac{|\mathrm{tr}(V_t)^2-\mathrm{tr}(V_t^2)|^2}{|\mathrm{tr}(V_t)|^4}.
    \label{eq:ratio_trace_ev}
\end{align} 
In the long-time limit, we assume that $|\lambda_{t,1}|\gg|\lambda_{t,2}|\gg|\lambda_{t,3}|$ is satisfied, and thus $\Omega_t^\lambda$ approaches $4|\lambda_{t,2}/\lambda_{t,1}|^2$.  
Here, we use the Cauchy-Schwarz inequality
\begin{align}
    \left(\overline{\sqrt{b_t}}\right)^2
    =\left(\overline{\sqrt{a_t}\sqrt{\frac{b_t}{a_t}}}\right)^2
    \leq\overline{a_t}\overline{\left(\frac{b_t}{a_t}\right)}
    \label{eq:Cauchy-Schwarz_inequality}
\end{align}
for arbitrary positive quantities $a_t$ and $b_t$. 
If we take $a_t=|\mathrm{tr}(V_t)^2-\mathrm{tr}(V_t^2)|^2$ and $b_t=|\mathrm{tr}(V_t)|^4$, Eq. (\ref{eq:Cauchy-Schwarz_inequality}) leads to
\begin{align}
    \left(\overline{\left|\mathrm{tr}(V_t)\right|^2}\right)^2\left/\:
    \overline{\left|\mathrm{tr}(V_t)^2-\mathrm{tr}(V_t^2)\right|^2}\right.
    \leq\overline{1/\Omega_t^\lambda}.
    \label{eq:inequality_omega}
\end{align}

Below, we evaluate the left-hand side of Eq. (\ref{eq:inequality_omega}). 
To this end, we use $\mathrm{tr}(A)\mathrm{tr}(B)=\mathrm{tr}(A \otimes B)$, where the trace in the right-hand side is carried out in $\mathsf{H}^{\otimes2}$ with $\mathsf{H}^{\otimes m}$ being the extended space whose dimension is $X^m$. 
Then, $\overline{|\mathrm{tr}(V_t)|^2}$ can be written as
\begin{align}
    \overline{|\mathrm{tr}(V_t)|^2}
    =\overline{\mathrm{tr}(V_t \otimes V_t^*)}=\mathrm{tr}\left[\mathcal{Q}^t\right],
    \label{eq:average_bt_root}
\end{align}
where $\mathcal{Q}=\overline{Q_t \otimes Q_t^*}$ is a time-independent matrix that acts on the extended space $\mathsf{H}^{\otimes2}$. 
The largest eigenvalue of $\mathcal{Q}$, which we write as $\mu$, determines the growth rate of $\overline{|\mathrm{tr}(V_t)|^2}$. 
In the same way, $\overline{|\mathrm{tr}(V_t)|^4}$ becomes
\begin{align}
    \overline{|\mathrm{tr}(V_t)|^4}
    =\overline{\mathrm{tr}\left[(V_t \otimes V_t)\otimes(V_t^* \otimes V_t^*)\right]}    =\mathrm{tr}\left[\mathbb{Q}^t\right],
    \label{eq:average_tsts}
\end{align}
where $\mathbb{Q}=\overline{(Q_t \otimes Q_t)\otimes(Q_t^* \otimes Q_t^*)}$ is a time-independent $X^4 \times X^4$ matrix and the trace in the right-hand side is taken in $\mathsf{H}^{\otimes4}$. 
For evaluating the other terms in $\overline{\left|\mathrm{tr}(V_t)^2-\mathrm{tr}(V_t^2)\right|^2}$, we use $\mathrm{tr}(AB)=\mathrm{tr}[(A \otimes B)\mathcal{S}]$ where $\mathcal{S}$ is the swap operator acting on $\mathsf{H}^{\otimes2}$. 
Thus, we obtain
\begin{align}
    \overline{\mathrm{tr}(V_t^2)\mathrm{tr}(V_t^2)^*}
    =\overline{\mathrm{tr}\left[\{(V_t \otimes V_t)\mathcal{S}\}
    \otimes\{(V_t^* \otimes V_t^*)\mathcal{S}\}\right]}
    =\mathrm{tr}\left[\mathbb{Q}^t (\mathcal{S} \otimes \mathcal{S})\right],
    \label{eq:average_stst}\\
    \overline{\mathrm{tr}(V_t)^2\mathrm{tr}(V_t^2)^*}
    =\overline{\mathrm{tr}\left[(V_t \otimes V_t)
    \otimes\{(V_t^* \otimes V_t^*)\mathcal{S}\}\right]}
    =\mathrm{tr}\left[\mathbb{Q}^t (I_{X^2} \otimes \mathcal{S})\right],
    \label{eq:average_tsst}\\
    \overline{\mathrm{tr}(V_t^2)[\mathrm{tr}(V_t)^2]^*}
    =\overline{\mathrm{tr}\left[\{(V_t \otimes V_t)\mathcal{S}\}\otimes(V_t^* \otimes V_t^*)\right]}
    =\mathrm{tr}\left[\mathbb{Q}^t (\mathcal{S} \otimes I_{X^2})\right],
    \label{eq:average_stts}
\end{align}
where $I_{X^2}$ is the identity matrix in $\mathsf{H}^{\otimes2}$. 
Equations (\ref{eq:average_tsts})-(\ref{eq:average_stts}) result in
\begin{align}
    \overline{|\mathrm{tr}(V_t)^2-\mathrm{tr}(V_t^2)|^2}    =\mathrm{tr}\left[\mathbb{Q}^t\mathbb{S}\right]
    \label{eq:average_at}
\end{align}
where $\mathbb{S}=\mathcal{S} \otimes \mathcal{S} + I_{X^2} \otimes I_{X^2} - \mathcal{S} \otimes I_{X^2} - I_{X^2} \otimes \mathcal{S}$. 
We notice that $\mathbb{S}$ and $\mathbb{Q}$ commute each other, which means that they have the same eigenmodes. 
The eigenvalues of $\mathbb{S}$, denoted as $\rho$, take $0$ or $4$ since eigenvalues of $\mathcal{S}$ are $\pm1$ and all ingredients of $\mathbb{S}$ such as $\mathcal{S}\otimes\mathcal{S}$ and $I_{X^2}\otimes\mathcal{S}$ commute with each other. 
Therefore, we focus only on eigenmodes with $\rho=4$ and neglect the other eigenmodes with $\rho=0$, which do not contribute to dynamics of $\overline{|\mathrm{tr}(V_t)^2-\mathrm{tr}(V_t^2)|^2}$. 
Then, the largest eigenvalue of $\mathbb{QS}/4$, which we write as $\nu$, determines the growth rate of $\overline{|\mathrm{tr}(V_t)^2-\mathrm{tr}(V_t^2)|^2}$ in the long run. 
Thus, dynamics of $[\overline{|\mathrm{tr}(V_t)|^2}]^2/\overline{|\mathrm{tr}(V_t)^2-\mathrm{tr}(V_t^2)|^2}$ is determined through $|\mu^2/\nu|^t/4$ in the long-time limit. 

We assume that non-unitary matrices $\{Q_t\}$ are determined through a random variable $R$ with $\overline{R}=0$ and that the system is characterized by the noise parameter $\beta\propto\sqrt{\overline{R^2}}$, where the overline represents the average over the random realizations of $R$. 
Here, $\beta$ is the strength of randomness, and the second moment of $R/\beta$ is independent of $\beta$. 
We note that $R$ corresponds to $\beta z_{\eta,t}$ in the case of the model studied in the main text. 

For small $\beta$, we can expand $Q_t$ as
\begin{align}
    Q_t \simeq A_t + \beta B_t+ \beta^2 C_t,
    \label{eq:Qt_expansion_1}
\end{align}
where $B_t$ and $C_t$ are random matrices. 
Here, $\overline{B_t}=0$ is satisfied because $B_t$ only includes the first-order terms of $R/\beta$. 
We also assume that $A_t$ is independent of $R$ and $A_t^\dagger=A_t^{-1}$ such that the system exhibits unitary dynamics if $\beta=0$. 
We note that other randomness independent of $R$ can be included in dynamics. 
For example, $\{A_t\}$ in Eq. (\ref{eq:Qt_expansion_1}) can be random unitary matrices determined through a random variable $R'$ that is independent of $R$. 
The average represented by the overline is over all kinds of randomness in noisy dynamics. 
Due to $\overline{B_t}=0$, Eq. (\ref{eq:Qt_expansion_1}) leads to
\begin{align}
    \mathcal{Q}\simeq \mathcal{E}+\beta^2\mathcal{F},\ \ 
    \mathbb{Q}\simeq \mathbb{E}+\beta^2\mathbb{F}, 
    \label{eq:Q_expansion_extended}
\end{align}
where $\mathcal{E}=\overline{A_t \otimes A_t^*}$,  $\mathbb{E}=\overline{(A_t \otimes A_t)\otimes(A_t \otimes A_t)^*}$, and 
\begin{align}
\mathcal{F}&=\overline{B_t \otimes B_t^*+A_t \otimes C_t^*+C_t \otimes A_t^*},
\label{eq:F_2}\\
\mathbb{F}&=\overline{(A_t \otimes A_t)\otimes(B_t \otimes B_t)^*}+\overline{(A_t \otimes A_t)\otimes(A_t \otimes C_t)^*}+\mathrm{their\ permutations\ regarding\ }A_t,B_t,C_t. 
\label{eq:F_4}
\end{align}
We argue that these Taylor expansions of averaged matrices $\mathcal{Q}$ and $\mathbb{Q}$ in Eq. (\ref{eq:Q_expansion_extended}) typically make subleading order terms of $\mu,\nu$ be $\beta^2$ in a wide range of noisy non-unitary dynamics. 
In addition, we assume that the leading terms of $|\nu|$ and $|\mu|$ are $1$, resulting in $\ln(|\sqrt{\nu}/\mu|)=0$ with $\beta=0$. 
This assumption is valid since $[\overline{|\mathrm{tr}(V_t)|^2}]^2$ and $\overline{|\mathrm{tr}(V_t)^2-\mathrm{tr}(V_t^2)|^2}$ should not decay in unitary dynamics. 
Thus, Eq. (\ref{eq:Q_expansion_extended}) suggests that the decay rate of $[\overline{|\mathrm{tr}(V_t)|^2}]^2/\overline{|\mathrm{tr}(V_t)^2-\mathrm{tr}(V_t^2)|^2}$ typically exhibits the $\beta^2$ dependence with respect to the noise parameter, which gives the lower bound of $\overline{|\lambda_{t,1}/\lambda_{t,2}|^2}$ in the long run through the Cauchy-Schwarz inequality in Eq. (\ref{eq:inequality_omega}). 
We also stress that the analysis explained above is applicable to the long-time regime as well as the short-time regime. 
Thus, $\tau_\Omega^\lambda$ defined through $f_t=t\ln(|\sqrt{\nu}/\mu|)$ gives the upper bound of $\tau_\lambda$. 
While there is a possibility that $\tau_\lambda$ decays faster than $\tau_\Omega^\lambda$, we believe that the $\tau_\Omega^\lambda$ and $\tau_\lambda$ typically exhibit the same power law $\tau\propto\beta^{-2}$, which is confirmed in the model studied in the main text. 

\subsection{Example}
\label{subsec:CS_specific}
We can actually show that $\mathcal{Q}$ and $\mathbb{Q}$ are written in the form of Eq. (\ref{eq:Q_expansion_extended}) regarding the model explored in the main text. 
In the model, the time-evolution matrix for one step is defined as
\begin{align}
    Q_t=G_tU
    \label{eq:Qt}
\end{align}
Here, the matrices $U$ and $G_t$ are defined as
\begin{align}
    U&=\bigoplus_\zeta
    \left[\begin{array}{cc}
    e^{i\theta_1}\cos(\theta_3)&-e^{i\theta_2}\sin(\theta_3)\\
    e^{-i\theta_2}\sin(\theta_3)&e^{-i\theta_1}\cos(\theta_3)
    \end{array}\right],
    \label{eq:U}\\
    G_t&=\bigoplus_\eta
    \left[\begin{array}{cc}
    \cosh\left(\beta z_{\eta,t}\right)&
    \sinh\left(\beta z_{\eta,t}\right)\\
    \sinh\left(\beta z_{\eta,t}\right)&
    \cosh\left(\beta z_{\eta,t}\right)
    \end{array}\right]=
    \bigoplus_\eta
    \exp\left( \beta z_{\eta,t} \sigma_1 \right),
    \label{eq:Gt}
\end{align}
where $z_{\eta,t}$ is chosen from the box distribution
\begin{align}
    z_{\eta,t}\in[-1/2,+1/2].
    \label{eq:random-range}
\end{align}
See Fig. 2 (a) in the main text. 
From Eqs. (\ref{eq:Qt})-(\ref{eq:random-range}), we can understand that this model satisfies the assumptions that $A_t=U$ is unitary and $\overline{B_t}=0$ with $B_t=(\bigoplus_\eta z_{\eta,t}\sigma_1)U$. 
Equations (\ref{eq:U}) and (\ref{eq:Gt}) lead to
\begin{align}
    \mathcal{Q}=\mathcal{GU},\ \ \mathbb{Q}=\mathbb{GU},
    \label{eq:Q_extended}
\end{align}
where $\mathcal{G}=\overline{G_t \otimes G_t}$, $\mathcal{U}=U \otimes U^*$, $\mathbb{G}=\overline{(G_t \otimes G_t)\otimes(G_t \otimes G_t)}$, and $\mathbb{U}=(U \otimes U)\otimes(U \otimes U)^*$.
In this model, we can write down $\mathcal{Q}$ and $\mathbb{Q}$ through calculating $\mathcal{G}$ and $\mathbb{G}$ explicitly, not relying on the perturbative expansion of $Q_t$ in Eq. (\ref{eq:Qt_expansion_1}). 
To this end, we decompose $G_t$ as
\begin{align}
    G_t=HG_t^\mathrm{d}H
    \label{eq:Gt_decomposition}
\end{align}
where $G_t^\mathrm{d}$ is a diagonal matrix
\begin{align}
    G_t^\mathrm{d}=
    \bigoplus_\eta
    \left(\begin{array}{cc}
     e^{+\beta z_{\eta,t}} & 0 \\
     0 & e^{-\beta z_{\eta,t}}
    \end{array}\right)
    \label{eq:Gt_diagonal}
\end{align}
and $H$ is the Hadamard matrix
\begin{align}
    H=\bigoplus_\eta
    \frac{1}{\sqrt{2}}
    \left(\begin{array}{cc}
     1 & 1 \\
     1 & -1
    \end{array}\right).
    \label{eq:Hadamard-matrix}
\end{align}
The non-unitary matrix in Eq. (\ref{eq:Gt_diagonal}) describes the dynamics of photons in optical networks exposed to loss effects and postselections \cite{mochizuki2023distinguishability}, and such dynamics of a photon has been observed experimentally \cite{xiao2017observation}. 
Equations (\ref{eq:Gt_decomposition}) and (\ref{eq:Gt_diagonal}) result in
\begin{align}
  \mathcal{G}=\mathcal{H}\mathcal{G}^\mathrm{d}\mathcal{H},\ \ 
    \mathbb{G}=\mathbb{H}\mathbb{G}^\mathrm{d}\mathbb{H},
    \label{eq:Gamma_decomposition}
\end{align}
with $\mathcal{G}^\mathrm{d}$ and $\mathbb{G}^\mathrm{d}$ being diagonal matrices where $\mathcal{H}=H \otimes H$ and $\mathbb{H}=\mathcal{H}\otimes\mathcal{H}$. 
The diagonal matrix $\mathcal{G}^mathrm{d}$ is written as
\begin{align}
    \mathcal{G}^\mathrm{d}=\bigoplus_{\eta_1,\eta_2}
    \tilde{\mathcal{G}}(\eta_1,\eta_2),
    \label{eq:Gamma-2_diagonal}
\end{align}
where $\tilde{\mathcal{G}}(\eta_1,\eta_2)$ is a $4\times4$ matrix. 
If $\eta_1\neq\eta_2$, $\tilde{\mathcal{G}}(\eta_1,\eta_2)$ becomes
\begin{align}
    \tilde{\mathcal{G}}(\eta_1,\eta_2)
    =g^2(\beta)I_4,
    \label{eq:Gamma-2_tilde_not-equal}
\end{align}
where $g(\beta)=\frac{\sinh(\beta/2)}{\beta/2}$ and $I_4$ is the $4\times4$ identity matrix. 
When $\eta_1=\eta_2$, $\tilde{\mathcal{G}}(\eta_1,\eta_1)$ becomes
\begin{align}
    \tilde{\mathcal{G}}(\eta_1,\eta_1)=
    \left[\begin{array}{cccc}
     g(2\beta) & 0 & 0 & 0 \\
     0 & 1 & 0 & 0 \\
     0 & 0 & 1 & 0 \\
     0 & 0 & 0 & g(2\beta)
    \end{array}\right].
    \label{eq:Gamma-2_tilde_equal}
\end{align}
The diagonal matrix $\mathbb{G}^\mathrm{d}$ is written as
\begin{align}
    \mathbb{G}^\mathrm{d}
    =\bigoplus_{\eta_1,\eta_2,\eta_3,\eta_4}
    \tilde{\mathbb{G}}(\eta_1,\eta_2,\eta_3,\eta_4),
    \label{eq:Gamma-4_diagonal}
\end{align}
where $\tilde{\mathbb{G}}(\eta_1,\eta_2,\eta_3,\eta_4)$ is a $16\times16$ matrix. 
When all of $\eta_i$ with $i=1,2,3,4$ take different values, $\tilde{\mathbb{G}}(\eta_1,\eta_2,\eta_3,\eta_4)$ becomes
\begin{align}
    \tilde{\mathbb{G}}(\eta_1,\eta_2,\eta_3,\eta_4)=[g(\beta)]^4I_{16}
    \label{eq:Gamma-4_tilde_not-equal}
\end{align}
where $I_{16}$ is the $16\times16$ identity matrix. 
When $\eta_1=\eta_2$, $\eta_2\neq\eta_3$, $\eta_2\neq\eta_4$, and $\eta_3\neq\eta_4$, $\tilde{\mathbb{G}}(\eta_1,\eta_1,\eta_3,\eta_4)$ becomes
\begin{align}
    \tilde{\mathbb{G}}(\eta_1,\eta_1,\eta_3,\eta_4)
    =\tilde{\mathcal{G}}(\eta_1,\eta_1)
    \otimes\tilde{\mathcal{G}}(\eta_3,\eta_4),
    \label{eq:Gamma-4_tilde_equal_12}
\end{align}
where $\tilde{\mathcal{G}}(\eta_1,\eta_1)$ and $\tilde{\mathcal{G}}(\eta_3,\eta_4)$ respectively correspond to Eqs. (\ref{eq:Gamma-2_tilde_equal}) and (\ref{eq:Gamma-2_tilde_not-equal}).  
We can obtain $\tilde{\mathbb{G}}(\eta_1,\eta_2,\eta_3,\eta_4)$ regarding other combinations such as $\eta_2=\eta_4$, through the corresponding permutation for the matrix elements of Eq. (\ref{eq:Gamma-4_tilde_equal_12}). 
When $\eta_1=\eta_2$, $\eta_2\neq\eta_3$, and $\eta_3=\eta_4$, $\tilde{\mathbb{G}}(\eta_1,\eta_1,\eta_3,\eta_3)$ becomes
\begin{align}
    \tilde{\mathbb{G}}(\eta_1,\eta_1,\eta_3,\eta_3)
    =\tilde{\mathcal{G}}(\eta_1,\eta_1)\otimes\tilde{\mathcal{G}}(\eta_3,\eta_3)
    \label{eq:Gamma-4_tilde_equal_12-34}
\end{align}
where $\tilde{\mathcal{G}}(\eta_1,\eta_1)$ corresponds to Eq. (\ref{eq:Gamma-2_tilde_equal}). 
We can obtain $\tilde{\mathbb{G}}(\eta_1,\eta_2,\eta_3,\eta_4)$ regarding the other combinations, such as $\eta_1=\eta_3$ and $\eta_2=\eta_4$, through the permutation for the matrix elements of Eq. (\ref{eq:Gamma-4_tilde_equal_12-34}). 
When $\eta_1=\eta_2=\eta_3$ and $\eta_3\neq\eta_4$, $\tilde{\mathbb{G}}(\eta_1,\eta_1,\eta_1,\eta_4)$ becomes
\begin{align}
    \tilde{\mathbb{G}}(\eta_1,\eta_1,\eta_1,\eta_4)=
    g(\beta)\left[\begin{array}{cccccccc}
     g(3\beta)I_2 & 0_2 & 0_2 & 0_2 & 0_2 & 0_2 & 0_2 & 0_2\\
      0_2 & g(\beta)I_2 & 0_2 & 0_2 & 0_2 & 0_2 & 0_2 & 0_2\\
     0_2 & 0_2 & g(\beta)I_2 & 0_2 & 0_2 & 0_2 & 0_2 & 0_2\\
     0_2 & 0_2 & 0_2 & g(\beta)I_2 & 0_2 & 0_2 & 0_2 & 0_2\\
     0_2 & 0_2 & 0_2 & 0_2 & g(\beta)I_2 & 0_2 & 0_2 & 0_2\\
     0_2 & 0_2 & 0_2 & 0_2 & 0_2 & g(\beta)I_2 & 0_2 & 0_2\\
     0_2 & 0_2 & 0_2 & 0_2 & 0_2 & 0_2 & g(\beta)I_2 & 0_2\\
     0_2 & 0_2 & 0_2 & 0_2 & 0_2 & 0_2 & 0_2 & g(3\beta)I_2
    \end{array}\right]
    \label{eq:Gamma-4_tilde_equal_123}
\end{align}
where $I_2$ and $0_2$ are the $2\times2$ identity matrix and the null matrix, respectively. 
We can obtain $\tilde{\mathbb{G}}(\eta_1,\eta_2,\eta_3,\eta_4)$ regarding the other combinations, such as $\eta_2=\eta_3=\eta_4$, through the corresponding permutation for the matrix elements of Eq. (\ref{eq:Gamma-4_tilde_equal_123}). 
When $\eta_1=\eta_2=\eta_3=\eta_4$, $\tilde{\mathbb{G}}(\eta_1,\eta_1,\eta_1,\eta_1)$ becomes
\begin{align}
    \tilde{\mathbb{G}}(\eta_1,\eta_1,\eta_1,\eta_1)=
    \left[\begin{array}{cccccccccccccccc}
     g(4\beta) & 0 & 0 & 0 & 0 & 0 & 0 & 0 & 0 & 0 & 0 & 0 & 0 & 0 & 0 & 0\\
     0 & g(2\beta) & 0 & 0 & 0 & 0 & 0 & 0 & 0 & 0 & 0 & 0 & 0 & 0 & 0 & 0\\
     0 & 0 & g(2\beta) & 0 & 0 & 0 & 0 & 0 & 0 & 0 & 0 & 0 & 0 & 0 & 0 & 0\\
     0 & 0 & 0 & 1 & 0 & 0 & 0 & 0 & 0 & 0 & 0 & 0 & 0 & 0 & 0 & 0\\
     0 & 0 & 0 & 0 & g(2\beta) & 0 & 0 & 0 & 0 & 0 & 0 & 0 & 0 & 0 & 0 & 0\\
     0 & 0 & 0 & 0 & 0 & 1 & 0 & 0 & 0 & 0 & 0 & 0 & 0 & 0 & 0 & 0\\
     0 & 0 & 0 & 0 & 0 & 0 & 1 & 0 & 0 & 0 & 0 & 0 & 0 & 0 & 0 & 0\\
     0 & 0 & 0 & 0 & 0 & 0 & 0 & g(2\beta) & 0 & 0 & 0 & 0 & 0 & 0 & 0 & 0\\
     0 & 0 & 0 & 0 & 0 & 0 & 0 & 0 & g(2\beta) & 0 & 0 & 0 & 0 & 0 & 0 & 0\\
     0 & 0 & 0 & 0 & 0 & 0 & 0 & 0 & 0 & 1 & 0 & 0 & 0 & 0 & 0 & 0\\
     0 & 0 & 0 & 0 & 0 & 0 & 0 & 0 & 0 & 0 & 1 & 0 & 0 & 0 & 0 & 0\\
     0 & 0 & 0 & 0 & 0 & 0 & 0 & 0 & 0 & 0 & 0 & g(2\beta) & 0 & 0 & 0 & 0\\
     0 & 0 & 0 & 0 & 0 & 0 & 0 & 0 & 0 & 0 & 0 & 0 & 1 & 0 & 0 & 0\\
     0 & 0 & 0 & 0 & 0 & 0 & 0 & 0 & 0 & 0 & 0 & 0 & 0 & g(2\beta) & 0 & 0\\
     0 & 0 & 0 & 0 & 0 & 0 & 0 & 0 & 0 & 0 & 0 & 0 & 0 & 0 & g(2\beta) & 0\\
     0 & 0 & 0 & 0 & 0 & 0 & 0 & 0 & 0 & 0 & 0 & 0 & 0 & 0 & 0 & g(4\beta)\\
    \end{array}\right].
    \label{eq:Gamma-4_tilde_equal_1234}
\end{align}
Since $g(\beta)\simeq1+\beta^2/24$ for small $\beta$, Eqs. (\ref{eq:Gamma-2_diagonal})-(\ref{eq:Gamma-4_tilde_equal_1234}) lead to Eq. (\ref{eq:Q_expansion_extended})
where $\mathcal{E}=\mathcal{U}$ and $\mathbb{E}=\mathbb{U}$. 
Here, $\mathcal{F}$ ($\mathbb{F}$) is a complicated matrix obtained from the Taylor expansion of $\mathcal{G}$ ($\mathbb{G}$) and the multiplication of $\mathcal{H}$ and $\mathcal{U}$ ($\mathbb{H}$ and $\mathbb{U}$).

\section{Perturbation analysis on the ratio of singular values}
\label{sec:perturbation}
As discussed in the main text, we can also evaluate the relaxation time $\tau_\Lambda$ through the singular values $\{\Lambda_{t,i}\}$ of $V_t$ in Eq. (\ref{eq:singular-values}),
where $\Lambda_{t,1}\geq\Lambda_{t,2}\cdots\geq\Lambda_{t,X}$. 
Here, we discuss the perturbation analysis for evaluating $\tau_\Lambda$.

If $\Lambda_{t,1}\gg\Lambda_{t,2}$ is satisfied, the fluctuating bunching state emerges. 
We perturbatively calculate
\begin{align}
    \Omega_t^\Lambda
    =\frac{\mathrm{tr}(\tilde{V}_t)^2-\mathrm{tr}(\tilde{V}_t^2)}{\mathrm{tr}(\tilde{V}_t)^2},
    \label{eq:ratio_trace_sv}
\end{align}
which approaches $2(\Lambda_{t,2}/\Lambda_{t,1})^2$ in the long run provided that  $\Lambda_{t,1}\gg\Lambda_{t,2}\gg\Lambda_{t,3}$ is satisfied. 
We again expand $Q_s$ as
\begin{align}
    Q_s \simeq A_s + \beta B_s+ \beta^2 C_s,
    \label{eq:Qt_expansion_2}
\end{align}
for small $\beta$, where $B_s$ and $C_s$ are random matrices. 
Remember that $\overline{B_s}=0$ and $A_s^\dagger=A_s^{-1}$ are assumed. 
We note that the analysis explained below is not restricted to the model explored in the main text and thus is applicable to a wide range of noisy non-unitary dynamics, as long as the assumptions above are satisfied. 
Up to the second-order terms of $\beta$, $V_t$ can be expanded as
\begin{align}
    V_t \simeq \mathscr{A}_{t,1} + 
    \beta\sum_s \mathscr{A}_{t,s+1} B_s \mathscr{A}_{s-1,1}+
    \beta^2\sum_s \mathscr{A}_{t,s+1}C_s \mathscr{A}_{s-1,1},
    \label{eq:Vt_expansion}
\end{align}
where $\mathscr{A}_{s_2,s_1}=A_{s_2}A_{s_2-1} \cdots A_{s_1+1}A_{s_1}$. 
For consistency of the notation, we define $\mathscr{A}_{0,1}=\mathscr{A}_{t,t+1}=I_X$ with $I_X$ being the $X \times X$ identity matrix. 
In Eq. (\ref{eq:Vt_expansion}), second order terms of $\beta$ like $\beta^2\mathscr{A}_{t,s_2+1}B_{s_2}\mathscr{A}_{s_2-1,s_1+1}B_{s_1}\mathscr{A}_{s_1-1,1}$ with $s_2-1>s_1+1$ are ignored. 
This is because we carry out the average over randomness at the end of the analysis, and such terms vanish due to $\overline{B_s}=0$ with $B_{s_1}$ and $B_{s_2}$ being independently distributed for $s_1 \neq s_2$. 
Following the same rule, $\tilde{V}_t$ becomes
\begin{align}
    \tilde{V}_t \simeq I_X + 
    \beta\sum_s \left(\mathscr{A}_{s,1}^\dagger B_s \mathscr{A}_{s-1,1}+\mathrm{h.c.}\right)+
    \beta^2\sum_s \left(\mathscr{A}_{s,1}^\dagger C_s \mathscr{A}_{s-1,1}+\mathrm{h.c.}\right)+
    \beta^2\sum_s \mathscr{A}_{s-1,1}^\dagger 
    B^\dagger_s B_s\mathscr{A}_{s-1,1}
    \label{eq:Vt_tilde_expansion}
\end{align}
where $I_X$ is the $X \times X$ identity matrix. 
With Eq. (\ref{eq:Vt_tilde_expansion}), $\mathrm{tr}\left(\tilde{V}_t\right)^2$ and $\mathrm{tr}\left(\tilde{V}_t^2\right)$ can be written as
\begin{align}
    \mathrm{tr}\left(\tilde{V}_t\right)^2&\simeq
    X^2+2X\beta\sum_s\mathrm{tr}(A^\dagger_s B_s+B_s^\dagger A_s)
    +\beta^2\sum_s[\mathrm{tr}(A^\dagger_s B_s+B^\dagger_s A_s)]^2
    +2X\mathrm{tr}(A^\dagger_s C_s+C^\dagger_s A_s+B^\dagger_s B_s)
    \label{eq:Vt_tilde_trace-square_expansion}\\
    \mathrm{tr}\left(\tilde{V}_t^2\right)&\simeq
     X+2\beta\sum_s\mathrm{tr}(A^\dagger_s B_s+B_s^\dagger A_s)
     +\beta^2\sum_s\mathrm{tr}[4B^\dagger_s B_s+(A^\dagger_s B_s)^2+(A_s B^\dagger_s)^2+2A^\dagger_s C_s+2A_s C^\dagger_s ]
    \label{eq:Vt_tilde_square-trace_expansion}
\end{align}
respectively. 
Substituting Eqs. (\ref{eq:Vt_tilde_trace-square_expansion}) and (\ref{eq:Vt_tilde_square-trace_expansion}) into Eq. (\ref{eq:ratio_trace_sv}), we can obtain
\begin{align}
    \Omega_t^\Lambda\simeq 1-\frac{1}{X}
    +\frac{\beta^2}{X^3}\sum_s[\mathrm{tr}(A^\dagger_s B_s+B^\dagger_s A_s)]^2
    -\frac{\beta^2}{X^2}\sum_s\mathrm{tr}[(A^\dagger_s B_s)^2+(B^\dagger_s A_s)^2]
    -\frac{2\beta^2}{X^2}\sum_s\mathrm{tr}(B^\dagger_s B_s).
    \label{eq:omega_expansion}
\end{align}
With large $X$, Eq. (\ref{eq:omega_expansion}) results in 
\begin{align}
   \overline{\ln\left(\Omega_t^\Lambda\right)}\simeq-\mathrm{tr}\left[2\overline{B^\dagger_s B_s}+\overline{(A^\dagger_s B_s)^2}+\overline{(B^\dagger_s A_s)^2}\right]\frac{\beta^2}{X^2}t.
    \label{eq:log-omega_average_expansion_supple}
\end{align}
We note that $\overline{(A^\dagger_s B_s)^2}$, $\overline{(B^\dagger_s A_s)^2}$, and $\overline{B^\dagger_s B_s}$ do not depend on $s$. 
Equation (\ref{eq:log-omega_average_expansion_supple}) is valid in the short-time regime for which $\beta^2t/X\ll1$ is satisfied, since $\mathrm{tr}\left[2\overline{B^\dagger_s B_s}+\overline{(A^\dagger_s B_s)^2}+\overline{(B^\dagger_s A_s)^2}\right]$ should be proportional to the system size $X$. 
However, the power law of the relaxation time $\tau_\Omega^\Lambda\propto\beta^{-2}$ obtained from Eq. (\ref{eq:log-omega_average_expansion_supple}) is consistent with actual relaxation time $\tau_\Lambda$ obtained through numerical simulations, which can be confirmed regarding two models explored in the main text and in Sec. \ref{sec:chen-model}.  
For the model explored in the main text, we have $A_s=U$, $B_s=Z_s U$, and $C_s=Z_s^2 U/2$, where $Z_s=\bigoplus_\eta z_{\eta,s}\sigma_1$ with $\sigma_1$ being one of the Pauli matrices. 
Then, the trace in Eq. (\ref{eq:log-omega_average_expansion_supple}) becomes $\mathrm{tr}\left[2\overline{B^\dagger_s B_s}+\overline{(A^\dagger_s B_s)^2}+\overline{(B^\dagger_s A_s)^2}\right]=4\mathrm{tr}\left(\overline{Z_s^2}\right)=X/3$, which leads to Eq. (\ref{eq:log-omega_average_expansion}).

\section{Results in a different model}
\label{sec:chen-model}
\begin{figure}[htbp]
\begin{center}
\includegraphics[width=17cm]{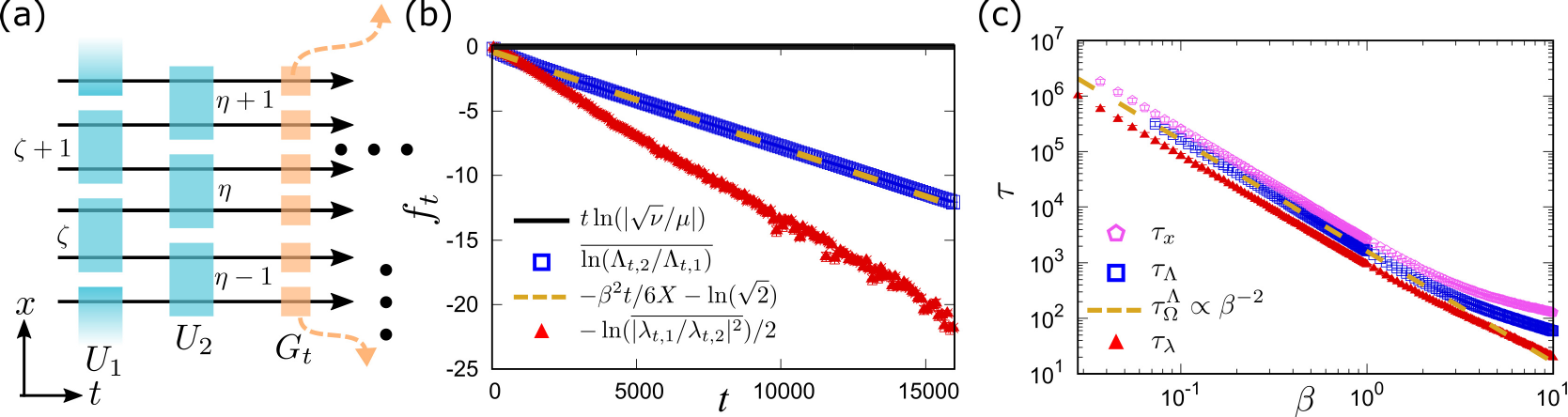}
\caption{Results of a model introduced in Sec. \ref{sec:chen-model}. 
(a) Schematic picture of the dynamics for one step. 
Blue rectangles and orange squares correspond to unitary dynamics by $U_1,\,U_2$, and non-unitary dynamics by $G_t$, respectively. 
(b) Ratios of $|\lambda_{t,i}|$ and $\Lambda_{t,i}$, which indicate that the fluctuating bunching state emerges after long-time dynamics. 
Red and blue symbols show $-\ln\left(\overline{|\lambda_{t,1}/\lambda_{t,2}|^2}\right)/2$ and $\overline{\ln\left(\Lambda_{t,2}/\Lambda_{t,1}\right)}$, respectively. 
The average is taken over $10^5$ samples. 
The orange broken line is $-\beta^2t/6X-\ln(\sqrt{2})$ which is derived through the perturbation analysis on $\overline{\ln\left(\Omega_t^\Lambda\right)}$. 
While this perturbation analysis is justified if $\beta^2t/X$ is small, as discussed in Sec. \ref{sec:perturbation}, we can see that its prediction turns out to be valid even for the long-time regime. 
The black line represents $t\ln\left(\left|\sqrt{\nu}/\mu\right|\right)$ obtained through numerical diagonalization of $\mathcal{Q}$ and $\mathbb{Q}$. 
Here, $\mu=\sqrt{\nu}$, and thus $t\ln\left(\left|\sqrt{\nu}/\mu\right|\right)$ is always zero, making the bound in Eq. (\ref{eq:inequality_omega}) useless. 
The noise parameter is $\beta=0.3$. 
(c) Relaxation times as functions of the noise parameter $\beta$ with $c=10^{-6}$, which exhibit the power law $\tau\propto\beta^{-2}$. 
Blue, red, and purple symbols respectively represent $\tau_\Lambda$, $\tau_\lambda$, and $\tau_x$ with $n=2$. 
The average in $f_t$ is taken over $10^2$ samples for obtaining one $\tau$.
We then reiterate obtaining $\tau$ for $10^2$ times and take the average over them. 
The orange broken line corresponds to relaxation times defined through the perturbative analysis on $\overline{\ln\left(\Lambda_{t,2}/\Lambda_{t,1}\right)}$.
Since this relaxation time is obtained by $-\beta^2\tau_\Omega^\Lambda/6X-\ln(\sqrt{2})=\ln(c)$, the orange broken line is exactly proportional to $\beta^{-2}$. 
In (b) and (c), the parameters are $(\theta_{11},\theta_{21},\theta_{31},\theta_{12},\theta_{22},\theta_{32})=(0.33\pi,0.41\pi,0.25\pi,0.22\pi,0.13\pi,0.25\pi)$, where the system size is $X=20$. 
}
\label{fig:chen-model}
\end{center}
\end{figure}

We show that the fluctuating bunching state and the power law of relaxation times are also observed in a model different from that in the main text. 
This suggests that these two phenomena are universal in a wide range of noisy non-unitary dynamics of bosons. 
The model which we consider in this section is described by $Q_t$ in Eq. (\ref{eq:Qt}) with $U$ and $G_t$ different from Eqs. (\ref{eq:U}) and (\ref{eq:Gt}). 
The time-independent unitary matrix $U$ is composed of two unitary matrices, $U=U_2U_1$, where
\begin{align}
    U_1&=\bigoplus_\zeta
    \left[\begin{array}{cc}
    e^{i\theta_{11}}\cos(\theta_{31})&-e^{i\theta_{21}}\sin(\theta_{31})\\
    e^{-i\theta_{21}}\sin(\theta_{31})&e^{-i\theta_{11}}\cos(\theta_{31})
    \end{array}\right],
    \label{eq:U1_chen-model}\\
    U_2&=\bigoplus_\eta
    \left[\begin{array}{cc}
    e^{i\theta_{12}}\cos(\theta_{32})&-e^{i\theta_{22}}\sin(\theta_{32})\\
    e^{-i\theta_{22}}\sin(\theta_{32})&e^{-i\theta_{12}}\cos(\theta_{32})
    \end{array}\right].
    \label{eq:U2_chen-model}
\end{align}
Here, $\zeta$ and $\eta$ correspond to two-site unit cells described in Fig. \ref{fig:chen-model} (a). 
 The non-unitary random matrix $G_t$ is defined as
\begin{align}
    G_t=\bigoplus_xe^{\beta z_{x,t}},
    \label{eq:Gt_chen-model}
\end{align}
where $z_{x,t}$ is sampled from the box distribution
\begin{align}
    z_{x,t}\in[-1/2,+1/2].
    \label{eq:random-range_chen-model}
\end{align}
If we take $\theta_{11}=\theta_{12}=0,\,\theta_{21}=\theta_{22}=\pi/2$, and $\theta_{31}=\theta_{32}=\pi/4$, this model corresponds to that analyzed in Ref. \cite{chen2021non}. 

Blue and red symbols in Fig. \ref{fig:chen-model} (b) show $\overline{\ln\left(\Lambda_{t,2}/\Lambda_{t,1}\right)}$ and $-\ln\left(\overline{|\lambda_{t,1}/\lambda_{t,2}|^2}\right)/2$, respectively, which indicate that the fluctuating bunching state emerges in the long-time regime. 
While the parameters $\{\theta_{ij}\}$ adopted in Fig. \ref{fig:chen-model} are different from those in Ref. \cite{chen2021non}, we have also confirmed the decay of $\overline{\ln\left(\Lambda_{t,2}/\Lambda_{t,1}\right)}$ and $-\ln\left(\overline{|\lambda_{t,1}/\lambda_{t,2}|^2}\right)/2$ with the same parameters as Ref. \cite{chen2021non}. 
This implies that the easiness for the boson sampling problem reported in Ref. \cite{chen2021non} originates from the fluctuating bunching state since classical computers can sample the probability distribution of bosons efficiently if the bunching state is realized \cite{mochizuki2023distinguishability}. 
The black line and orange broken line are $\ln\left(|\sqrt{\nu}/\mu|^t\right)$ and $-\beta^2t/6X-\ln(\sqrt{2})$ obtained by analyses in Secs. \ref{sec:Cauchy-Schwarz_inequality} and \ref{sec:perturbation}, respectively. 
Note that $\sqrt{\nu}=\mu$ in the present model, which indicates that the Cauchy-Schwarz inequality discussed in Sec. \ref{sec:Cauchy-Schwarz_inequality} does not provide a good bound for the relaxation time $\tau_\Omega^\lambda$.
Figure \ref{fig:chen-model} (c) shows relaxation times $\tau_\lambda$ (red symbols),  $\tau_\Lambda$ (blue symbols), and $\tau_x$ (purple symbols). 
We can understand that all relaxation times are proportional to $\beta^{-2}$ for small $\beta$; the three relaxation times are parallel to the orange broken line, $6X|\ln(c)+\ln(\sqrt{2})|\beta^{-2}$. 

\end{document}